# Multiferroic spin-superfluid and spin-supersolid phases in MnCr$_2$S$_4$

Alexander Ruff,[1] Zhaosheng Wang,[2] Sergei Zherlitsyn,[2] Joachim Wosnitza,[2] Stephan Krohns,[1] Hans-Albrecht Krug von Nidda,[1] Peter Lunkenheimer,[1] Vladimir Tsurkan,[1,3] Alois Loidl[1*]

[1]Experimental Physics V, Center for Electronic Correlations and Magnetism,
University of Augsburg, 86159 Augsburg, Germany
[2]Hochfeld-Magnetlabor Dresden (HLD-EMFL), Helmholtz-Zentrum Dresden-Rossendorf,
01328 Dresden, Germany
[3]Institute of Applied Physics, MD 2028, Chisinau, R. Moldova

*Corresponding author. Email: alois.loidl@physik.uni-augsburg.de (A.L.)

Spin supersolids and spin superfluids reveal complex canted spin structures with independent order of longitudinal and transverse spin components. This work addresses the question whether these exotic phases can exhibit spin-driven ferroelectricity. Here we report the results of dielectric and pyrocurrent measurements of MnCr$_2$S$_4$ as function of temperature and magnetic field up to 60 T. This sulfide chromium spinel exhibits a Yafet-Kittel type canted spin structure at low temperatures. As function of external magnetic field, the manganese spins undergo a sequence of ordering patterns of the transverse and longitudinal spin components, which can be mapped onto phases as predicted by lattice-gas models including solid, liquid, super-fluid, and supersolid phases. By detailed dielectric and pyrocurrent measurements, we document a zoo of multiferroic phases with sizable ferroelectric polarization strongly varying from phase to phase. Using lattice-gas terminology, the title compound reveals multiferroic spin-superfluid and spin-supersolid phases, while the antiferromagnetic solid is paraelectric.

## INTRODUCTION

This manuscript aims to combine two fascinating topics of recent solid-state research, namely multiferroicity and supersolidity. The discovery of spin-driven ferroelectricity by Kimura et al. (*1*) revived the large field of multiferroicity (*2*), which is of great fundamental and technological importance (*3,4*). It is nowadays well established that various spin configurations and exchange mechanisms can induce ferroelectric (FE) ground states and result in complex magnetoelectric ($H,T$) phase diagrams (*5*). The claim of the appearance of supersolidity in liquid $^4$He (*6*), a hybrid between a solid and a superfluid, raised considerable scientific interest, later was discarded (*7*), but stimulated enormous further theoretical and experimental work. For example, supersolid phases were recently identified in ultra-cold atomic gases (*8,9*) and, moreover, magnetic analogs of supersolids, so called spin supersolids, were found in some spinel compounds (*10,11,12*). These magnetic variants of supersolids and superfluids are characterized by complex spin structures, with independent order of the longitudinal and transverse spin components. Non-collinear spin arrangements often give rise to multiferroicity and, thus, it seems natural to combine these topical areas of research and to search for FE spin supersolids and superfluids. This is the primary motivation of the present work.

More than four decades ago, controversial reports on the space group of spinel minerals were discussed in the scientific literature. Conventional cubic spinels usually are assigned to the space group Fd$\bar{3}$m. However, a number of physical properties were seemingly better understood assuming the loss of inversion symmetry giving rise to the space group F$\bar{4}$3m (*13,14,15*). This proposal was further elucidated by Schmid and Ascher (*16*), showing that a system with space group F$\bar{4}$3m undergoing a ferrimagnetic phase transition will feature magnetic phases that permit piezoelectricity, piezomagnetism, significant magnetoelectric effects, and even can exhibit ferroelectricity. To our knowledge, in the eighties this controversy fell asleep. However, triggered by the discovery of spin-driven multiferroicity in perovskite manganites (*1*), magnetoelectric effects including polar order have been reported in a number of spinels (*17,18,19,20*). In this class of compounds, only the occurrence of multiferroicity in $CoCr_2O_4$ (*19*) could be understood in terms of a canonical spin-driven mechanism, where the occurrence of ferroelectricity is explained either by a spin-current model (*21*) or by utilizing an inverse Dzyaloshinskii-Moriya (DM) interaction (*22,23*).

$MnCr_2S_4$ is a normal spinel compound, which undergoes ferrimagnetic ordering at 65 K and exhibits a triangular Yafet-Kittel (YK) type (*24*) canted spin structure below 5 K (*25,26*). From the very beginning, the question concerning the magnetic ground state of $MnCr_2S_4$, revealing either a collinear or canted spin-structure, was controversially debated (*27*) and finally was solved by Plumier and Sougi (*28*). The existing experimental evidence of a non-collinear canted ground state is consistent with $F\bar{4}3m$ symmetry (*29*), being potentially magnetoelectric and multiferroic (*16*). In addition, the title compound gained considerable attention due to the fact that it exhibits a complex (*H,T*) phase diagram, with an ultra-robust magnetization plateau around 40 T and a sequence of ordered spin structures with separate order parameters of longitudinal and transverse spin components (*30,31,12*). Most interestingly, these complex spin structures can be mapped onto theoretical predictions of phases obtained from quantum lattice-gas models (*32,33*), including magnetic analogs of solid, liquid, superfluid, and supersolid phases (*12*).

In this manuscript, by performing dielectric and pyrocurrent measurements as function of temperature and external magnetic field, we prove that the low-temperature YK phase of $MnCr_2S_4$ is FE. Moreover, by mapping the complex spin structures onto predictions of quantum lattice-gas models, we provide striking experimental evidence that - in the (*H,T*) phase diagram up to 60 T - the liquid and solid phases are paraelectric (PE), while the superfluid and supersolid phases are FE.

**RESULTS**

We start the discussion concerning the possibility of multiferroic spin-supersolid and spin-superfluid phases in $MnCr_2S_4$ with a description of its complex spin structures as function of temperature and magnetic field. The non-trivial non-collinear ground state results from dominating ferromagnetic (FM) Cr-Cr and competing antiferromagnetic (AFM) Cr-Mn and Mn-Mn exchange. In the normal spinel structure, the manganese ions are located on a bipartite diamond lattice. Neglecting the exchange between these sublattices, both Mn spins would be oriented strictly antiparallel to the chromium moments. However, weak AFM exchange between the sublattices finally cants the manganese spins, resulting in the YK ground state. In the spinel structure, Mn-Mn exchange is weak, as it is always mediated via intermediate S-Cr-S ion bridges (*34*). The exchange paths between the two manganese sublattices, Mn(1) and Mn(2), are schematically indicated in Fig. 1A. Due to mainly 180° bonds, the paths 1-2-3 and 4-2-5 are the dominant AFM exchange paths (*34*). The chromium spins are aligned parallel to the external magnetic field, which in Fig. 1A is along the crystallographic [111] direction. Figure 1B shows the bipartite manganese diamond lattice, which consists of two interpenetrating fcc lattices. Only the manganese spins with their canted spin arrangement, arising from the AFM exchange between the two sublattices, are shown. In this figure, it is assumed that the external magnetic field and, hence, the chromium moments, point along the [111] direction, the body diagonal of the cube. Finally, Fig. 1C documents the evolution of the spin order as function of an external magnetic field as proposed in (*12*). On increasing field, the YK ground state, which is stable at low external magnetic fields, is followed by the intermediate phase 1 (IM1), by the ideal antiferromagnet (AFM), by the intermediate phase 2 (IM2), by the inverse YK (IYK) phase, and finally by the ferromagnetic spin-polarized phase (FM). The IYK and FM phases, predicted by theory, were not experimentally detected due to the limited high-field range. As will be discussed later, using symmetry arguments for the longitudinal and transverse manganese spin components, the YK and IYK phases correspond to spin superfluids, while both intermediate phases resemble spin-supersolid phases (*32,33*). The question that we would like to answer in this work is, whether some of these complex spin arrangements can induce ferroelectricity and, hence, lead to multiferroic phases.

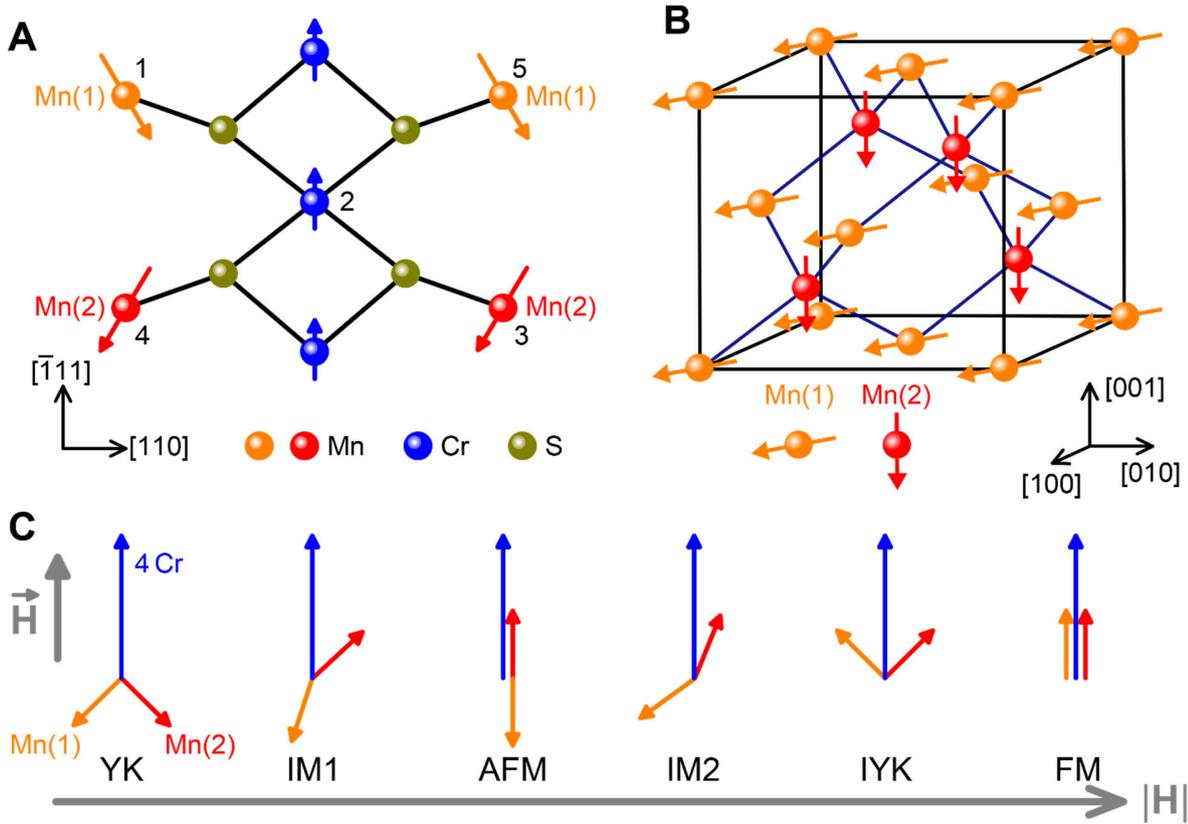

**Fig. 1. Spin arrangements in MnCr₂S₄.** (**A**) Exchange paths between the two manganese sublattices, Mn(1) and Mn(2) of the bipartite diamond lattice. Chromium, manganese, and sulphur ions, including the localized spins at the manganese and chromium sites in the Yafet-Kittel phase, are indicated. Chromium spins are aligned in the external magnetic field along [111]. Via exchange paths of the predominantly 180° bonding angles (1-2-3 or 4-2-5), the two manganese sites are antiferromagnetically coupled resulting in a YK structure in low external magnetic fields $H$ and at low temperatures. (**B**) The two manganese sites of the diamond lattice within the unit cell of the normal spinel structure. The chromium spins (not shown) are aligned along the crystallographic [111] direction (body diagonal of the fcc lattice). The shown spin canting of the manganese spins corresponds to Yafet-Kittel order at low magnetic fields. (**C**) Schematic spin order as function of an external field $H$: On increasing fields, Yafet-Kittel (YK) order is followed by an intermediate phase 1 (IM1), by antiferromagnetic (AFM) order, by an intermediate phase 2 (IM2), by an inverse YK (IYK) order, and finally, by the field-polarized ferromagnetic order (FM). The intermediate phases correspond to spin supersolidity because of the coexistence of transverse and longitudinal AFM spin order of the two manganese sites. In (A) and (B) the magnetic moments are not drawn to scale. In (C) the moments are drawn to scale, taking the magnetization per unit cell into account.

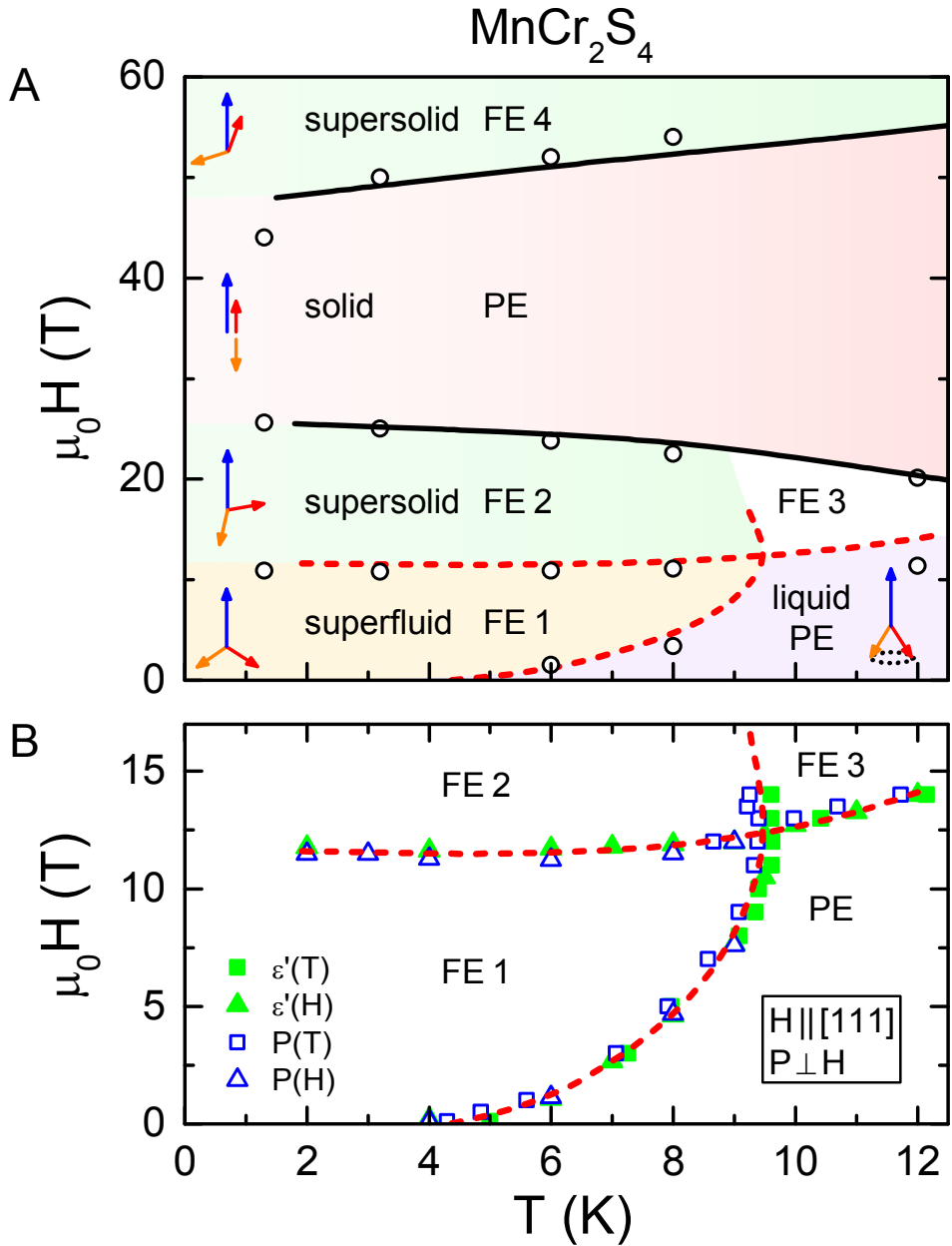

**Fig. 2. (*H,T*) phase diagram of MnCr₂S₄.** (**A**) Schematic phase diagram as determined in (*12*). The phase boundaries as deduced in these experiments are indicated by solid lines. Anomalies resulting from the present high-field pyrocurrent experiments are shown by empty circles. Spin configurations [arrows: chromium (blue), manganese (red and orange)] and the polar order (PE, FE) determined in the present work are indicated. Four ferroelectric phases (FE1, FE2, FE3, and FE4) with different polarizations were identified. Dashed lines were derived from the present dielectric and pyrocurrent experiments, revealing the additional phase FE3. Following predictions of quantum lattice-gas models, the phases are assigned analogous to liquid, superfluid, supersolid, and solid states (*32,33*). (**B**) Low-field phases as determined from the temperature and field dependence of the dielectric constant $\varepsilon'$ and polarization *P* up to 14 T. Dashed lines are drawn to guide the eye. Paraelectric (PE) and ferroelectric phases FE1, FE2, and FE3 are indicated.

A solid starting point for the interpretation of the temperature and field dependence of the dielectric and pyrocurrent results, is the complex (*H,T*) phase diagram of MnCr₂S₄. It was derived earlier by magnetization and ultrasonic experiments up to 60 T (*12*). Figure 2A documents this phase diagram, with the phase boundaries as previously reported (*12*) indicated by solid lines. The circles indicate data points derived from the present high-field pyrocurrent measurements, which will be discussed below. For the phases shown in Fig. 2A, the chromium spins are always aligned parallel to the external field. With respect to this quantization axis, the manganese spins reveal order of the transverse as well as the longitudinal spin components, i.e., perpendicular or parallel to the chromium spins, respectively. Only the spin order of the two manganese sublattices is important for the sequence of magnetic phases in MnCr$_2$S$_4$.

In low external magnetic fields, the YK structure is the stable spin configuration with symmetrically canted spins and a FM component exactly antiparallel to the chromium spins (cf. Fig. 1). As was detected by magnetization experiments (*12,31*), the canting angle continuously increases up to approximately 11 T. Here the system undergoes a meta-magnetic phase transition where both canted manganese spins turn with respect to the chromium moments and the external field, establishing AFM order of the transverse as well as of the longitudinal spin components, as is illustrated in Fig. 1C. This spin-configuration is indicated as intermediate phase IM1 and corresponds to a spin-supersolid phase. Above 25 T, the manganese spins form ideal antiferromagnetic order, with both subsystems being aligned parallel or antiparallel to the chromium moments and, hence, to the external magnetic field (AFM state in Fig. 1C). This regime extends to 50 T and is characterized by the detected robust magnetization plateau, which is solely determined by the magnetization of the completely field-polarized chromium moments (*12*). Beyond 50 T, the canted manganese spins further turn with respect to the chromium moments, again with AFM order of both, longitudinal and transverse spin components. This supersolid spin structure again is the stable spin configuration, but now with the net longitudinal moment aligned parallel to the external field (cf. Fig. 1C). Remarkably, when viewed from the manganese spin configurations, the spin structures seem to be symmetric with respect to an external magnetic field of 40 T. Here the external field exactly compensates the chromium exchange at the manganese site, locally creating an effective zero field. At low magnetic fields, the YK phase is only stable up to 5 K and, on increasing temperature, is followed by a partly paramagnetic (PM) phase with ferromagnetically aligned chromium spins, ferrimagnetic order of the longitudinal manganese spin components, but disordered transverse spin components.

As mentioned above, the spin configurations of MnCr$_2$S$_4$ as function of an external magnetic field can be directly compared to predictions of quantum lattice-gas models (*32,33*). In these models, FM and PM states correspond to liquid phases, AFM transverse spin order represents the superfluid phase, coexistent transverse and longitudinal AFM spin order characterizes the supersolid phase, while AFM order of the longitudinal spin components only, denotes the solid phase. This sequence of phases is indicated in Fig. 2A. Again, it should be noted that these phases appear mirrored for positive and negative effective magnetic fields, with zero field corresponding to an external magnetic field of 40 T. While it is certainly intriguing that, 45 years after their theoretical prediction, spin structures that are symmetry analogs to complex states of matter have been experimentally verified (*12*), it seems even more exciting to check whether these spin structures as shown in Fig. 1C also induce polar states and drive these magnetic phases towards multiferroicity.

Figure 3 shows the low-temperature dielectric constant $\varepsilon'$ of MnCr$_2$S$_4$ as function of temperature and magnetic field. In these experiments, the external magnetic field was parallel to the crystallographic [111] direction, with the electric field oriented perpendicular to the magnetic field. Figure 3A documents the temperature dependence of $\varepsilon'$ between 2 and 15 K for a series of external magnetic fields between 0 and 14 T. At 0 T, the dielectric constant shows an abrupt step-like increase below ~ 5 K, which is of order 3 % of the absolute values. The temperature of this dielectric anomaly exactly corresponds to the transition from the ferrimagnetic phase, with disordered transverse spin components of the manganese spins (*26*), to the planar triangular YK spin structure, where the spins of the two manganese sites of the diamond lattice are antiferromagnetically coupled to the chromium spins, but are slightly canted. This significant dielectric anomaly provides a first hint that indeed FE order occurs in the YK phase, i.e., that this spin analog of a superfluid phase is multiferroic. Such anomalies are characteristic features of improper ferroelectrics, where FE order is driven by another order parameter, and were observed in many spin-driven multiferroics (*1,35*). With increasing external magnetic field, the anomaly shifts to higher temperatures and follows the phase boundary of the YK phase (see Figs. 2A and B). At the same time, the step-like enhancement of the dielectric constant becomes considerably smaller. The origin of this decrease of the dielectric anomaly when crossing the YK phase boundary for finite fields, partly remains an open question. One possible interpretation is that, in the absence of an external magnetic field, the sample is in a multidomain state, while in finite fields a single-domain state, with all chromium moments aligned parallel to the external field, is reached. The significant decrease of the dielectric anomaly with increasing magnetic fields implies that the probing electric field is not strictly in line with the macroscopic polarization.

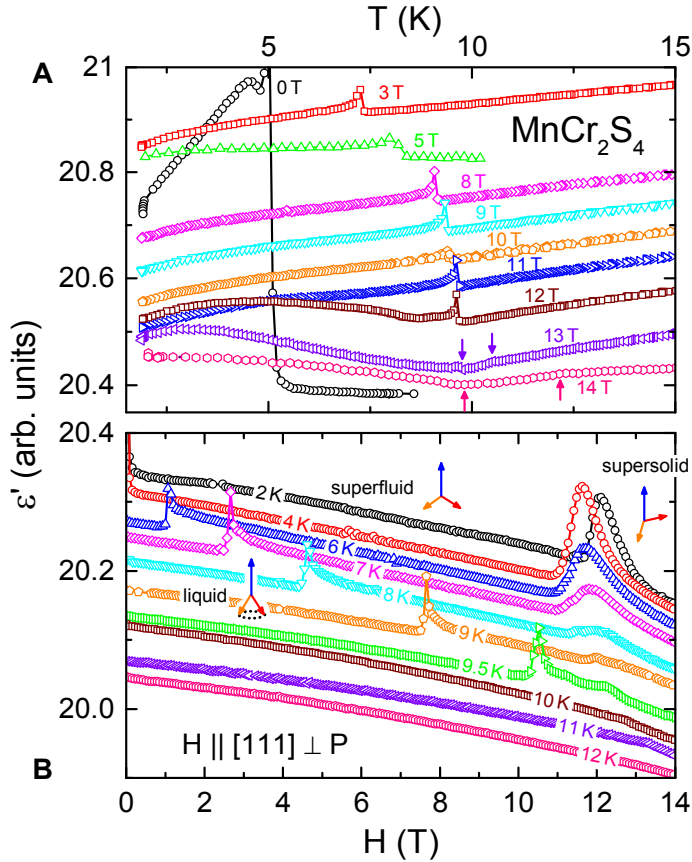

**Fig. 3. Temperature and magnetic-field dependence of the dielectric constant $\varepsilon'$ in MnCr$_2$S$_4$.** These measurements were performed at a frequency of 1 kHz with the external magnetic field $H$ parallel to the crystallographic [111] direction and the probing electric field perpendicular to the magnetic field. **(A)** Temperature dependence of $\varepsilon'(T)$ from 2 to 15 K at a series of external magnetic fields between 0 and 14 T. Anomalies at 13 and 14 T are indicated by arrows. **(B)** Magnetic-field dependence $\varepsilon'(H)$ from 0 to 14 T at a series of temperatures from 2 to 12 K. For temperatures $T \geq 10$ K, the characteristic peak-like anomalies vanish and are replaced by a change of slope of $\varepsilon'(H)$. The spin patterns (chromium: blue; manganese: orange and red) are indicated and denoted analogous to phases derived in lattice-gas models. The minor differences in the absolute values of the dielectric constants in (A) and (B) correspond to experimental uncertainties in the field- and temperature-dependent measuring cycles.

Here, we focus on the critical temperatures of the observed anomalies. On increasing magnetic fields up to 11 T, the dielectric anomaly shifts up to 9 K, with a very similar overall shape. At 12 T it remains close to 9 K, but an additional broad hump evolves at low temperatures, signaling continuous changes in the low-temperature phase. Interestingly, on further increasing fields, the shape of the dielectric anomaly changes considerably [see $\varepsilon'(T)$ in Fig. 3A at 12 T as compared to the results at 13 and 14 T]. On decreasing temperatures and at the two highest fields, a minor change of slope of the temperature-dependent dielectric constant (at about 11 K for 13 T and 12 K for 14 T; right arrows in Fig. 3A) is followed by a minimum (left arrows) and the mentioned subsequent hump in the low-temperature phase. These findings already indicate that a further phase with different polar states may exist for higher fields and temperatures, which will be elucidated below.

Figure 3B documents the magnetic-field dependence of $\varepsilon'$ between 0 and 14 T, for a series of temperatures between 2 and 12 K. A high-field peak close to 12 T is followed by a low-field anomaly, which is strongly field dependent. The latter, e.g., appears close to 1T at 6 K and approaches zero fields at the lowest temperatures. According to the phase diagram documented in Fig. 2A, the state between these two dielectric anomalies corresponds to the YK phase, which strongly narrows with increasing temperature and finally vanishes above about 10 K. The two dielectric anomalies certainly correspond to polar phase transitions: the low-field anomaly indicates the transition from the PE high-temperature phase with PM disorder of the transverse spin components to a FE phase with YK spin order. As indicated in Figs. 1C and 2A, on increasing magnetic fields the YK phase is followed by a spin configuration where the triangular configuration of the manganese spins partly rotates into the direction of the external field (*12,31*). The anomaly close to 12 T revealed by Fig. 3B signals that at this magnetic phase boundary the polar order also changes. However, based on the $\varepsilon'$ data alone, it is unclear if this high-field anomaly signals a transition into a PE or into a FE phase with different polarization. At 8 K, the YK phase still exists between about 5 and 12 T. In contrast, at 10 K the lower phase transition is absent, but a change of slope can be identified in $\varepsilon'(H)$ close to 13 T. Here the high-field phase seems to have a different polar configuration, in accord with the hints at the existence of an additional phase provided by the $\varepsilon'(T)$ data of Fig. 3A, discussed in the previous paragraph.

To determine the polar ground state of this sequence of magnetic phases, we performed detailed pyro- and magneto-current measurements. From the results, we determined the FE polarization and its direction, depending on magnitude and sign of the pyrocurrent. Figure 4A shows the temperature-dependent polarization $P(T)$ as function of the external magnetic field. In these experiments, the magnetic field was oriented along the crystallographic [111] direction and the pyrocurrent was measured perpendicular to the magnetic field. We made a series of additional experiments measuring the ferroelectric polarization also parallel to the external magnetic field. We found identical ordering temperatures and a similar size of the ferroelectric polarization, signaling that the polarization in the title compound is neither strictly perpendicular nor parallel to the crystallographic [111] direction. This fact will be discussed later, also including the high-field results.

Already a first look at Fig. 4A reveals significant polarization at low temperatures for almost all magnetic fields. This finding clearly proves the polar nature of the covered magnetic phases, not only including the spin-superfluid YK phase, but also the spin-supersolid phase with transverse and longitudinal Mn-spin components (IM1 in Fig. 1C). At zero external field, the polarization at low temperatures is enhanced by a factor of more than two compared to non-zero fields, corresponding to the significantly larger dielectric anomaly when crossing the phase boundary revealed by Fig. 3A. For finite magnetic fields, the low-temperature polarization remains rather constant and extends to successively higher temperatures as the field increases. The temperature-dependent scan at 12 T results in a polarization that is close to zero, crosses a FE phase between 8 and 10 K, and is zero again for higher temperatures. Most interestingly, for magnetic fields > 12 T the polarization changes sign and reveals a clear two-step behavior. As documented in Fig. 2A, below 12 T these scans mainly probe the YK phase, which is ferroelectric (FE1) and followed by a PE phase at higher temperatures. However, at low temperatures, above 12 T the YK phase is followed by a second FE phase (FE2) with opposite polarization, the spin-supersolid IM1 phase. In addition, even a third FE phase (FE3) seems to exist in a narrow range of external magnetic fields above 12 T and at temperatures between 10 and 12 K. At magnetic fields > 1 T we expect single-domain magnetic order, with the chromium spins aligned parallel to the external field and the longitudinal component of the canted manganese spins antiparallel to the field. Summarizing the complex field and temperature dependence of the pyrocurrent, we conclude that both the spin-superfluid YK and the spin-supersolid IM1 phases are multiferroic with coexistent magnetic and FE order, and that $MnCr_2S_4$ reveals a zoo of multiferroic phases at low temperatures and magnetic fields below 14 T.

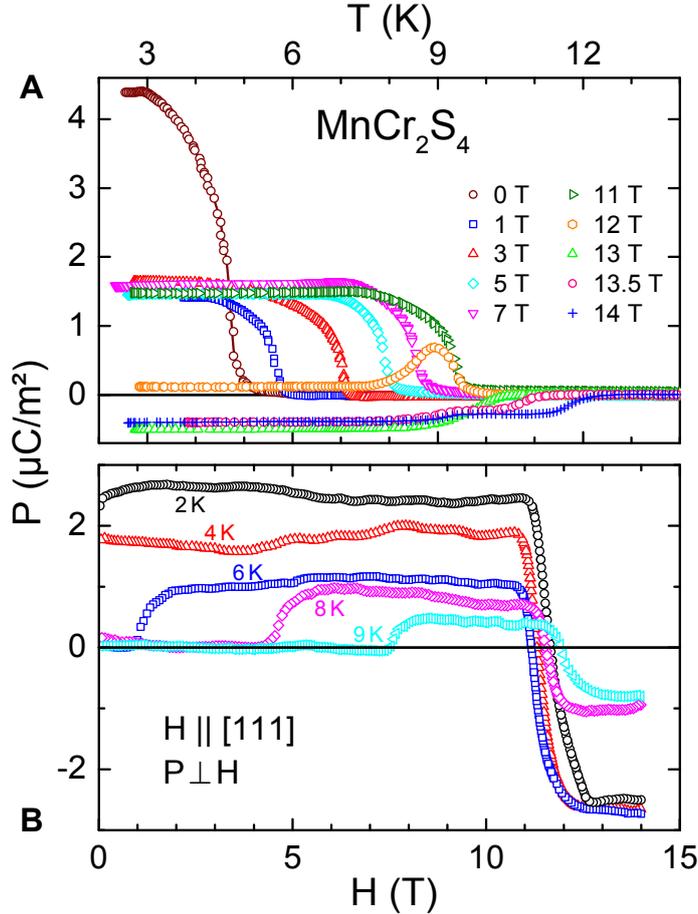

**Fig. 4. Temperature and magnetic-field dependence of the electric polarization *P* in $MnCr_2S_4$.** Measurements were performed with the external magnetic field parallel to the crystallographic [111] direction and with the pyro- or magnetocurrent measured perpendicular to the magnetic field. **(A)** Temperature dependence of the polarization for a series of external magnetic fields between 0 and 14 T. Negative polarization signals switched FE order. The two-step behavior for fields ≥ 13 T indicates subsequent FE phases with different polarization. **(B)** Magnetic-field dependence of the polarization for a series of temperatures between 2 and 9 K. The magnetic-field range of the YK phase is strongly temperature dependent and followed by a FE phase with switched polarization. Above 5 K, the low-field phase is PE. At 8 K, the YK phase is stable approximately between 5 and 12 T.

Further evidence for this scenario is provided by Fig. 4B, documenting the magnetic-field dependent polarization in $MnCr_2S_4$ from 0 to 14 T as measured at a sequence of temperatures between 2 and 9 K. These experiments again clearly document that the YK phase is ferroelectric (FE1) and that the stability range of the YK phase shrinks with increasing temperature. In low fields, the title compound is multiferroic only below 5 K. In good accord with the results of the $\varepsilon'(T,H)$ measurements (Fig. 3), the low-field boundary of this multiferroic phase increases with increasing field and it is stable, e.g., between 7 and 12 T at 9 K. At all investigated temperatures, the FE polarization $P(H)$ switches at magnetic fields above about 12 T and changes sign. This characterizes the transition from the YK to the IM1 magnetic phases. As documented in Fig. 4B, the latter phase also is ferroelectric (FE2), well confirming our conclusions from the *P(T)* results of Fig. 4A.

Figure 4B also reveals a decrease of the polarization plateaus with increasing temperature, both for fields below as well as above the phase boundary close to 12 T. When comparing Figs. 4A and B, we observe some discrepancies in the absolute values of the polarization. The values of the low-field polarization are of order 1 - 2 $\mu C/m^2$ in the field scans, while they are rather constant and of order 1.5 $\mu C/m^2$ in the temperature scans in a field range from 1 to 11 T. Larger polarization differences between the temperature and field scans are encountered when comparing the values of the high-field phase. It is not yet clear if this slight differences of the absolute values of the polarization result from experimental uncertainties, from different history dependences or different prepoling conditions of the samples in the different runs.

The thorough dielectric and pyroelectric measurements documented in Figs. 3 and 4 allow the construction of a detailed low-temperature phase diagram below 14 T (Fig. 2B). The observed polar phases can now be correlated with the magnetic phase diagram documented in Fig. 2A. At low fields and elevated temperatures, the manganese spins form a spin structure, with the longitudinal component antiparallel aligned to the chromium moments, but with the transverse component being still disordered (cf. Fig. 1C and arrows in Fig. 2A) (*26*). This phase is PE. As described above, on decreasing temperature, the manganese spins undergo YK order. This phase is FE with positive polarization and is assigned as FE1. On increasing magnetic fields, the canting angle of the YK spin structure opens and at 12 T, MnCr$_2$S$_4$ undergoes a metamagnetic transition where the canted manganese spins instantaneously turn with respect to the external field. Now, the transverse as well as the longitudinal components of the manganese spins are antiferromagnetically ordered. The phase with this spin configuration is also ferroelectric (FE2), however, with the polarization now switched to negative values, when compared to the YK phase. Interestingly, on increasing temperature, this ferroelectric phase FE2 is followed by another, so far unknown phase (FE3). It also is ferroelectric, but with reduced polarization, which probably points along the same crystallographic direction as for phase FE2. The spin order of this multiferroic phase still has to be determined. Note the appearance of a tetracritical point close to 10 K and 12 T, which will be discussed later.

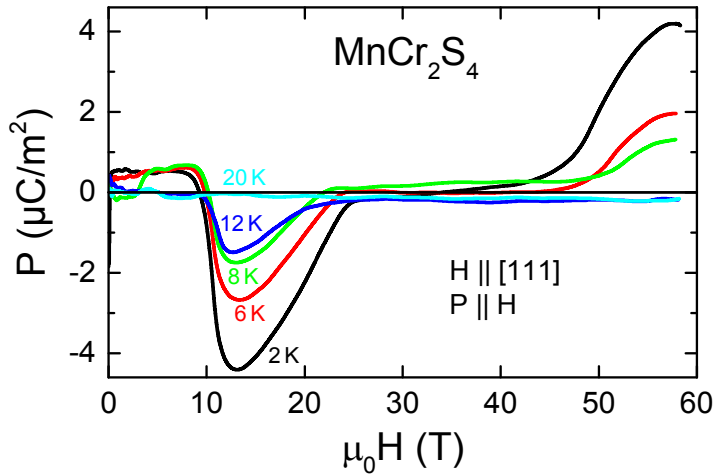

**Fig. 5. Magnetic-field dependence of the electric polarization *P* in MnCr$_2$S$_4$ up to 60 T.** These measurements were performed for a series of temperatures between 2 and 20 K with the external magnetic field parallel to the crystallographic [111] direction and with the pyrocurrent measured parallel to the magnetic field.

The dielectric and pyrocurrent experiments on MnCr$_2$S$_4$, documented in Figs. 3 and 4 and summarized in Fig. 2B, were complemented by high-field pyrocurrent experiments up to 60 T. These experiments have been performed with the magnetic field along the crystallographic [111] direction. In documenting the high-field experiments we focus on the results obtained with the pyrocurrent measured parallel to the external field. The resulting polarization $P(H)$ is shown in Fig. 5. At the lowest temperatures (2, 6, and 8 K) a finite positive polarization in the YK phase at low fields is followed by a negative polarization beyond 12 T, in good agreement with the findings of Fig. 4B. As stated above, close to 12 T the title compound enters into the FE2 phase discussed above, where the triangular spin structure becomes rotated with respect to the external field (phase IM1 in Fig. 1C). In this phase, the FE polarization switches with respect to the YK phase. As evidenced by the high-field results of Fig. 5, the polarization in this phase is not field-independent, but continuously decreases and finally becomes zero for fields close to 25 T. Here the chromium moments are ferromagnetically aligned in the external field and the manganese spins form ideal AFM order, with both sublattices parallel to the external field (AFM in Fig. 1C) (*12*). The continuous polarization decay in the IM1 phase results from the opening of the canting angle of the manganese spins with increasing fields, approaching a collinear spin structure. Ideal AFM order obviously finally results in zero polarization. Throughout the regime up to 50 T, characterized by ideal AFM spin order of the manganese moments, the polarization essentially remains zero. Beyond this $P = 0$ plateau region, when entering the IM2 spin structure [Fig. 1(C)], the polarization increases, showing positive values opposite to the polarization in phase IM1. At this phase boundary, the system enters into the FE phase FE4. In the phase IM2, the manganese spins again reveal AFM order of both spin components, but with the resulting FM components now pointing parallel to the external field and, hence, parallel to the chromium moments (IM2 in Fig. 1C). The continuous increase of the polarization signals the continuous evolution of the supersolid spin structure out of the ideal AFM order of the spin solid phase. At 12 K, $P(H)$ documents the occurrence of the phase FE3 with negative polarization, while the FE4 phase has shifted out of the field range of the present experiments. At 20 K the title compound is PE at all values of the external field. The anomalies found in the $P(H)$ curves of Fig. 5 are indicated by open circles in Fig. 2A, revealing a reasonable match with the known magnetic phase boundaries (*12*).

At this point, we would like to make some comments concerning the possible direction of the ferroelectric polarization in the multiferroic phases of MnCr$_2$S$_4$. Figs. 4B and 5 document experimental results of the polarization measured parallel and perpendicular to the crystallographic [111] direction, respectively. In additional low-field experiments (< 14 T), we observed identical phase boundaries in both directions with slightly different FE polarizations. In all experimental settings, we observed non-zero polarization and identical polarization switching from positive to negative, when crossing the superfluid to supersolid phase boundary. We conclude that in both phases the FE polarization can neither be parallel nor perpendicular to the [111] direction. At 2 K, for $P \perp [111]$ the polarization switches from ~ 2 to ~ −2 µC/m$^2$ (Fig. 4B), while for $P \parallel [111]$ it changes from ~ 1 to ~ −4 µC/m$^2$ (Fig. 5). Despite the fact that we have to consider some experimental uncertainties, it is clear that the polarization direction changes at the phase boundary. Moreover, the polarization seems to come closer parallel to the [111] direction in the supersolid phase. However, much more experimental work is needed to determine the polarization direction of all four identified FE phases.

## DISCUSSION

Based on our results on the temperature and field dependence of the dielectric constant and polarization, documented in Figs. 3 - 5, we were able to determine the polar order of the phases shown in Fig. 2A. The resulting states, either FE or PE, are indicated in Fig. 2A. We have identified four FE phases and two different PE phases. As discussed above, the spin structures of the manganese spins in these phases can be correlated with the phases as derived from lattice-gas models. At low temperatures and on increasing external magnetic fields, there is a sequence of spin superfluid, spin supersolid, and solid phases, followed by the mirrored sequence of supersolid, superfluid, and liquid phases on further increasing fields. The latter two phases are theoretically predicted and still have to be verified experimentally. As mentioned above, the phase diagram is symmetric about 40 T, where the manganese spins reside in zero effective magnetic field. The liquid or fluid phase corresponds either to a paramagnetic (at low magnetic fields) or to a fully spin-polarized ferromagnetic state (at high fields, not covered in the present work). Superfluid phases are characterized by triangular YK-type spin structures, with ordered transverse spin components. Supersolid phases correspond to spin configurations where the canted manganese spins turn with respect to the chromium moments and the external magnetic field. These canted arrangements of the manganese spins exhibit coexisting antiferromagnetic order of transverse and longitudinal spin components. The dielectric experiments performed in the course of this work allow the polar characterization of all these magnetic phases.

As mentioned above, it is clear that both the superfluid and the supersolid phases are FE. These phases are characterized by canted spin structures. At low fields, the canonical YK phase is FE with positive polarization. The superfluid phase at high fields, with inverse YK structure, could not be reached in the present pyrocurrent experiments, but probably is also FE. Both supersolid phases, characterized by concomitant transverse and longitudinal spin order are also FE. The low-field supersolid phase has negative polarization when compared to the YK phase. In addition, the two supersolid phases show opposite FE polarization (see Fig. 5). The extended region where the chromium moments are aligned with the external field and with the manganese spins showing ideal antiferromagnetic order is PE. It thus seems clear that only canted spin structures show FE polarization. At low fields, with increasing temperature the YK phase is followed by a magnetic phase with FM chromium order, predominantly ferrimagnetic alignment of the manganese spins, but with the transverse spin components being disordered. This phase also is PE, making the proposed spin structure of Ref. (*26*) very plausible. The phase FE3 is a new phase, which has been detected in the course of this work. The spin structure of this phase still remains to be to determined. As documented in Fig. 2B and mentioned above, a tetracritical point appears in the ($H,T$) phase diagram of $MnCr_2S_4$ close to 10 K and 12 T (Fig. 2B). At first glance, this seems to violate Gibbs' phase rule. However, with the appearance of supersolid phases in between superfluid and solid phases, tetracritical points have been predicted by lattice gas models (*33*) and were critically discussed (*36*).

Finally, we speculate about a possible spin-driven mechanism responsible for the fascinating multiferroic phase diagram of $MnCr_2S_4$. From Fig. 2 it seems clear that in the title compound a spin-driven mechanism must be responsible for the observed polar order and routes to ferroelectricity via charge order or via independent polar and spin systems certainly can be ruled out. Spin-driven ferroelectricity can occur via a spin-current mechanism (*21*) or, alternatively, via an inverse DM effect (*22,23*). However, *p-d* hybridization (*37,38*) or an exchange-striction mechanism (*5*) have also to be considered. In crystals with special symmetry, the lattice modulation by exchange striction can break the inversion symmetry and make the crystal polar (*5*). In the present case where ferroelectricity only appears concomitantly with canted non-collinear spin structures, we think that the exchange-striction mechanism can be excluded. *p-d* hybridization relies on spin-orbit coupling. The manganese ions with a half-filled *d* shell have vanishing spin-orbit coupling, also ruling out this mechanism. We think that the spin-current or inverse DM mechanisms seem to be the most promising routes for explaining the present results. The switching of FE polarization between the two supersolid phases IM1 and IM2 (see Fig. 5) provides further support of the DM mechanism as the vector chirality $S_i \times S_j$ changes sign in these two phases. However, there is no simple argument to explain the switching of FE polarization between superfluid and supersolid phases [Figs. 4(B) and 5]. Obviously, much more theoretical work is needed to explain the complexity of this zoo of multiferroic phases.

## MATERIALS AND METHODS

MnCr$_2$S$_4$ single crystals were grown by a chemical transport-reaction method from the ternary polycrystalline material prepared by solid-state reaction. A detailed structural, magnetic, and thermodynamic characterization of these crystals can be found in Refs. (*26,39*). The crystals had typical sizes of 3 × 3 × 2 mm$^3$. The majority of dielectric and pyrocurrent measurements was performed with the magnetic field along the crystallographic [111] direction and the probing electric field parallel or perpendicular to the magnetic field. Additional measurements with the magnetic field along [100] and [110] and the electric field parallel to the external magnetic field yielded similar results, with identical transition temperatures, identical switching behavior, and variations of the FE polarization never exceeding a factor of three. The dielectric measurements were performed with silver-paint contacts either in sandwich geometry or in a cap-like fashion, covering the opposite ends of the sample. The complex dielectric constants were measured for frequencies between 100 Hz and 10 kHz using an Andeen-Hagerling AH2700A high-precision capacitance bridge. For measurements between 1.5 and 300 K and in external magnetic fields up to 14 T, an Oxford Cryomagnet was used. To probe FE polarization, we measured both, the pyrocurrent at fixed magnetic field and the magnetocurrent at fixed temperature using a high-precision electrometer. To align the FE domains, on cooling electric poling fields of the order 1 kV/cm were applied. Electric polarization at high fields was measured by a pyroelectric technique (*40*). In these experiments pyrocurrent was mainly measured along the [111] direction, supplemented by some experiments along [110]. In all experiments, the current was measured parallel to the magnetic field. The pyrocurrent was captured through the voltage variation in a shunt resistor connected in series with the measurement circuit by a digital oscilloscope Yokogawa DL750 with a high sampling rate of 1 MS/s and a 16-bit resolution. Then, the polarization was calculated by integrating the pyrocurrent numerically. The sample was cooled with a poling field of about 1.0 kV/cm to the lowest temperature. After each measurement, the sample was heated up to 40 K and cooled down again with the poling field switched on.

**References and Notes**

**Acknowledgments**

**Funding:**
This research was supported by the Deutsche Forschungsgemeinschaft (DFG) via the Transregional Collaborative Research Center TRR 80: From Electronic Correlations to Functionality (Augsburg-Munich) and via the Research Center SFB 1143: From Frustration to Topology (Dresden). We acknowledge also support by HLD at HZDR, member of the European Magnetic Field Laboratory (EMFL).

**Author contributions:** A.R. and S.K. performed dielectric and pyrocurrent measurements up to 14 T. Z.W. and S.Z. designed and carried out the high-field experiments, supervised by J. W. Sample growth and characterization was performed by V.T. and H.-A. KvN. H.-A. KvN, P.L., V.T., and A.L. prepared the manuscript with input from all co-authors.

**Competing interests:** All authors declare that they have no competing interests.